# Why can a hydrophilic polyelectrolyte precipitate and redissolve below the critical micelle concentration of an oppositely-charged surfactant?

Huaisong Yong[1,2,3]*

[1]Department of Molecules & Materials, MESA+ Institute, University of Twente, 7500 AE Enschede, The Netherlands

[2]Institute Theory of Polymers, Leibniz-Institut für Polymerforschung Dresden e.V., D-01069 Dresden, Germany

[3]School of New Energy and Materials, Southwest Petroleum University, 610500 Chengdu, China

*Correspondences:  h.yong@utwente.nl (H. Y.); yonghuaisong@gmail.com (H. Y.)

**Abstract**

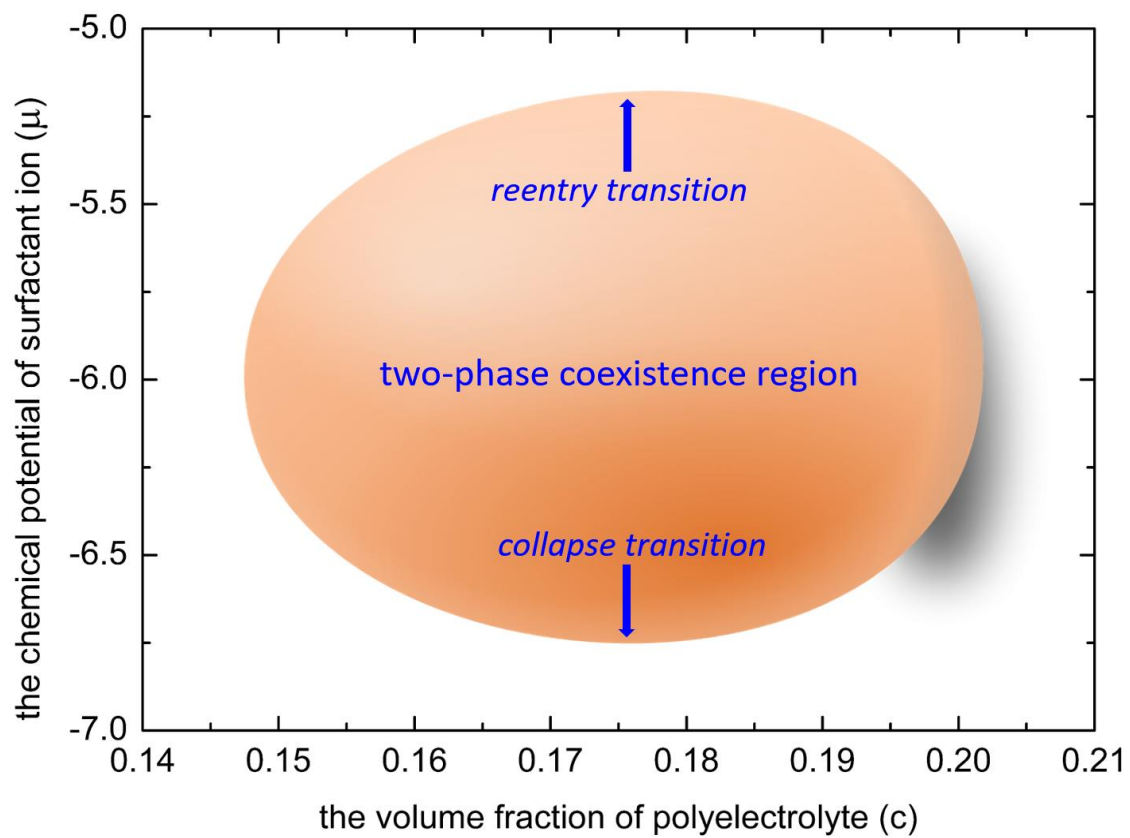

We theoretically study the reentrant condensation of a polyelectrolytes in the presence of an oppositely-charged surfactant, a phenomenon whose phase-transition mechanism remains under discussion. We focus on the adsorption and attraction effects of surfactant near/on polymer chains, and ignore their own non-essential mixing effects if surfactant molecules are far away from polymer chains. This approach allows us to construct a simple mean-field theory and solve it analytically, and finally rationalize the essential features (such as the "egg shape" of spinodal phase diagrams) of the reentrant condensation of a polyelectrolyte induced by diluted oppositely-charged surfactants. By theoretical analysis, we found that a strong electrostatic adsorption between the ionic monomers and surfactant ions is critical to understand the peculiar phenomenon that both the collapse and reentry transitions of polyelectrolytes can occur when the concentration of surfactant is lower than its bulk critical micelle concentration (CMC). The analytical solution of the theory indicates that a minimum coupling energy for the nonlinear hydrophobic-aggregation effect of adsorbed surfactant is essential for phase transition to occur, which explains why polyelectrolytes show phase transition only if the surfactant chain length is longer than a minimum length. The obtained results will shed light on a deep understanding of liquid-liquid phase separation in biological systems where ionic surfactant-like proteins/peptides bound to bio-polyelectrolytes play an important role.

## 1. Introduction

In this paper, we theoretically study the reentrant condensation of a polyelectrolyte in the diluted aqueous solution of an oppositely-charged surfactant, as sketched in **Figure 1a**, which has attracted significant attention in the past decades **[1-8]**. A remarkable feature of the reentrant condensation is that both the collapse and the reentry branches occur at rather low surfactant concentrations **[9-12]**, where the overall concentration of added surfactant (such as hexadecyl trimethylammonium bromides) is usually in the order of 5 mmol/L and below the bulk critical micelle concentration (CMC) of the surfactant. This feature implies that the effect of electrostatic screening actually plays a trivial role, since the Debye length of electrostatic screening is comparable with the size of an isolated polyelectrolyte chain or even larger in this case **[13]**. Another unusual feature of the reentrant condensation is that: The electrostatic binding between the charged monomer and oppositely-charged surfactant leads to the formation of insoluble polymer-surfactant coacervates if above a surfactant concentration, which is usually called the critical aggregation concentration (CAC). The CAC can be several orders of magnitude below the CMC of the surfactant **[14]**, which implies that the influence of the bulk micellar behaviors of the surfactant is negligible for the reentrant condensation. We note that in the literature [11, 12] a surfactant concentration is usually termed as the second critical aggregation concentration (CAC$_2$), at which the insoluble polymer-surfactant coacervates start to redissolve (i.e., the reentry branch of the polyelectrolyte condensation in **Figure 1a**).

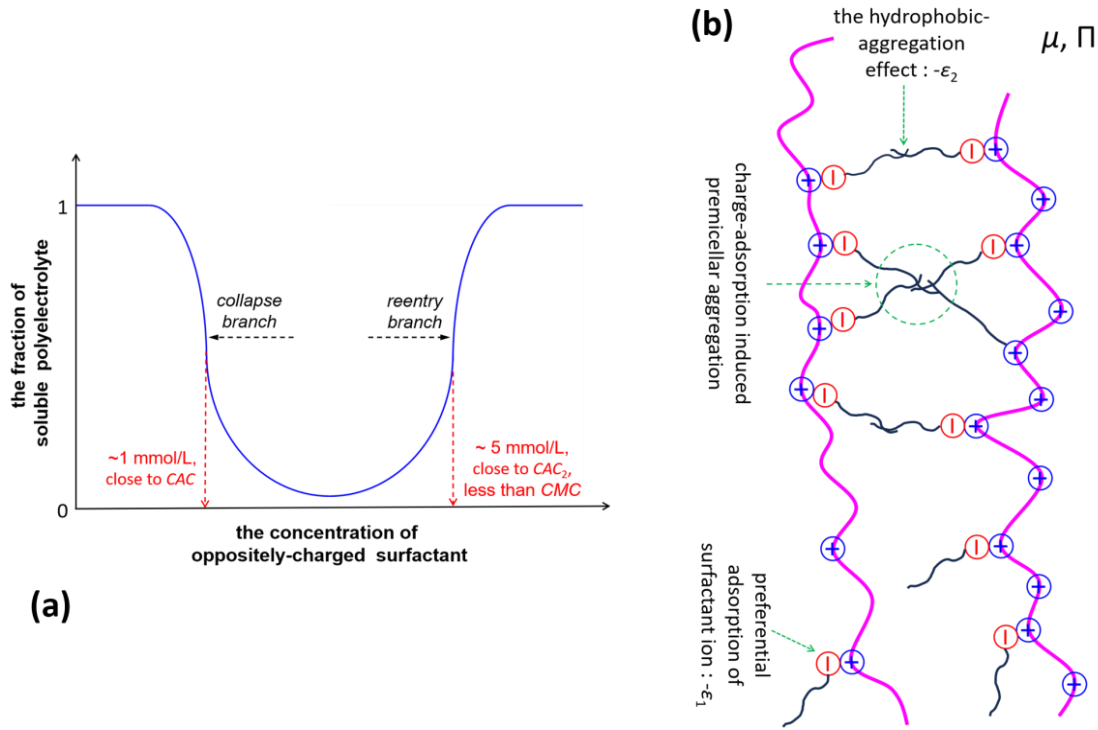

**Figure 1. (a)** A sketch description of the reentrant condensation of polyelectrolytes induced by diluted oppositely-charged surfactant. In the figure, the typical values of surfactant concentration for collapse and reentry transitions are quoted from **refs. [9-12]**. Notice that the reentrant condensation is not necessary to be symmetric with respect to surfactant concentrations. **(b)** A sketch description of the well-accepted phase-transition mechanism of electrostatic-adsorption induced hydrophobic aggregation [12, 15-18] for the collapse branch of the reentrant condensation as illustrated in **Panel a**. The ionic monomers (the symbol "⊕" in the figure) are preferentially adsorbed by oppositely-charged surfactant ions (the symbol "⊖" in the figure), and polyelectrolyte chains further form temporary bridges due to the hydrophobic-aggregation effect of adsorbed surfactant ions (contacted black lines in the figure). In the figure, the pink lines represent polymer chains and we do not show counterions of both polyelectrolytes and surfactants for simplification.

Clarifying the function of oppositely-charged surfactants in such reentrant condensation (as sketched in **Figure 1a**) is critical for a deep understanding of biological phase separations if ionic surfactant-like proteins/peptides bound to bio-polyelectrolytes play an important role [19-27], for example, for a better understanding of the DNA length on the co-condensation behaviors of proteins with DNAs [19, 25]. Also from the aspect of applied research, the elucidation of the phase-transition mechanism of the reentrant condensation is pivotal for developing new polymer formulations such as for drug delivery[3] and new materials such as porous polymer materials [28].

A well-accepted and general phase-transition mechanism [12, 15-18] for the collapse branch of the reentrant condensation (as sketched in **Figure 1a**) is that the surfactant ion replaces the polyelectrolyte counterions and preferentially adsorbs on the ionic monomers as well as forms electrostatic dipoles. This further results in a nonlinear hydrophobic-aggregation effect among the long hydrophobic tails of adsorbed surfactant ions. As illustrated by **Figure 1b**, this hydrophobic-aggregation effect among ionic monomers indicates that when the surfactant concentration reaches a critical value, the charge-adsorption induced premicellar aggregation will occur which will ultimately lead to the formation of insoluble polymer-surfactant coacervates. In contrast, the phase-transition mechanism for the reentry branch as sketched in **Figure 1a**, can be strongly complicated by the specific chemistry details of uncharged moieties of polyelectrolyte. It was reported that the hydrophobicity of uncharged monomers of polyelectrolyte affects and deceases the surfactant concentration that is necessary to trigger the reentry transition [12, 29].

Related to the nonlinear attraction effect among ionic monomers deriving from the electrostatic adsorption of surfactant [12, 15-18] (as illustrated by **Figure 1b**), theoretical rationalizations in the past decades such as chemical reaction-like theories [30, 31] and lattice/field-like theories [10, 18, 30, 32-42] have significantly promoted the understanding of phase behaviors of polyelectrolyte in the presence of oppositely-charged surfactant, such as recent interesting on the complex micellar structures of polymer-surfactant coacervates [31-34]. However, there remain two fundamental problems [43] to be addressed clearly for the reentrant condensation as sketched in **Figure 1a:** The first problem is why polyelectrolytes show phase transition only if the surfactant chain length is longer than a minimum length [44-47]. The second problem is why both the collapse and reentry transitions of polyelectrolytes can occur when the concentration of surfactant is lower than its bulk critical micelle concentration (CMC).

The primary goal of this work is to try to resolve the above two problems by exploiting the well-accepted phase-transition mechanism as sketched in **Figure 1b**. Compared with previous theoretical formalisms [10, 18, 30-42], in this work we focus on the adsorption and attraction effects of surfactant near/on polymer chains, and ignore their own non-essential mixing effects if surfactant molecules are far away from polymer chains. This approach allows us to construct a simple mean-field theory and solve it analytically for the reentrant condensation (as illustrated by **Figure 1a**). We found that a strong electrostatic adsorption between the ionic monomers and surfactant ions is critical to understand the peculiar phenomenon that both the collapse and reentry transitions of polyelectrolytes can occur when the concentration of surfactant is lower than its bulk critical micelle concentration (CMC). The analytical solution of the theory indicates that a minimum coupling energy for the nonlinear hydrophobic-aggregation effect of adsorbed surfactant is essential for phase transition to occur, which explains why polyelectrolytes show phase transition only if the surfactant chain length is longer than a minimum length.

In the rest of this article, a mean-field theory for the polyelectrolyte solution in terms of the free energy will be constructed in **Section 2**. Its analytical solution will be considered in detail in **Section 3**, where we will outline some general results of the mean-field theory and a simplified phase diagram of the polyelectrolyte solution will be discussed in this section. The applicability of the mean-field theory will be also discussed in this section. Final concluding remarks will be made in **Section 4**.

## 2. Methods and Model

In this section, we construct the mean-field theory for the polyelectrolyte solution in presence of an oppositely-charged surfactant based upon the Gibbs free energy and clarify its physics foundation.

As sketched in **Figure 1b**, we consider flexible polyelectrolytes with monovalent ionic monomers and monovalent counterions in aqueous solution. We denote $N$ as the number of monomers on a polyelectrolyte chain and $a$ as the size of each monomer along the direction of the polymer backbone. The charged monomers are distributed randomly on the polyelectrolyte chain and their fraction is denoted by $p$. For simplicity but without losing generality, we assume that the size of the ionized monomer, the size of counterion of polyelectrolyte, the size of counterion of surfactant, and the size of solvent molecules are the same. The overall volume fraction of monovalent ionic monomers and neutral monomers is denoted by $c$, then the volume fraction of monovalent counterions is $pc$. We denote the volume fraction of counterion of surfactant as $c_x$, then the volume fraction of surfactant ion is $nc_x$ where $n$ denotes the surfactant chain size in unit of a polyelectrolyte monomer.

Here and in the following, let's consider the free energy per unit of volume for an incompressible system if not otherwise noted specifically. The volume unit is given by the size of the solvent molecules in the spirit of the classical Flory–Huggins lattice model **[48, 49]**. We separate the system into two parts; the polymer chains with their enclosed solvents and small ions, and the bulk without polymer chains. The contribution from the mixing of polymer with solvent and added surfactant is given by $G_{sol}$, see the first line of **Equation (1)**. The bulk is just described by its osmotic pressure ($\Pi$) acting on the first part. If the volume of the polymer coils changes, then there is the mechanical work against the external pressures involved in the free energy change. This will become important when we minimize the free energy with respect to the monomer concentration ($c$). This approach indicates that the current formalism is under the framework of the isothermal–isobaric ensemble (*NPT* ensemble). Here and in the following, we consider energies in units of $k_B T$ where $k_B$ is the Boltzmann constant and $T$ is the thermodynamic temperature.

$$\begin{aligned}
G_{sol} &= \frac{c}{N}\ln(c) + pc\ln(pc) + c_x \ln(c_x) + nc_x \ln(nc_x) \\
&\quad + \left[1-(1+p)c-(1+n)c_x\right]\ln\left[1-(1+p)c-(1+n)c_x\right] + \Pi \\
G_{FH} &= \left[1-(1+p)c-(1+n)c_x\right]\left[p\varepsilon_{FH,1}+(1-p)\varepsilon_{FH,2}\right]c \\
\frac{G_{ads}}{pc} &= \frac{\varphi}{\lambda}\ln(\varphi) + (1-\varphi)\ln(1-\varphi) - \mu\varphi - \varepsilon_1 \frac{\varphi}{\lambda} + \chi_s \varphi(1-\varphi) \\
\frac{G_{attr}}{pc} &= -\varepsilon_2 \varphi^2 (1-\rho)(pc) \approx -\varepsilon_2 \varphi^2 (1-\varphi)(pc) \\
G_{DS} &\simeq -\frac{(\kappa a)^3}{8\pi(1+\kappa a)} \approx -\frac{(\kappa a)^3}{8\pi} = -\sqrt{\pi}\left(\frac{l_B}{a}\right)^{\frac{3}{2}}\left[pc+(1+n)c_x\right]^{\frac{3}{2}} \\
G(\varphi,c,c_x) &= G_{sol} + G_{FH} + G_{ads} + G_{attr} + G_{DS}
\end{aligned} \qquad (1)$$

The energy of non-electrostatic excluded-volume interactions between solvent (water) and the charge neutral part of monomers is given by the classical Flory-Huggins formalism **[48, 49]** as $G_{FH}$, see the second line of **Equation (1)**. The Flory-Huggins parameters between solvent and charged monomers is denoted by $\varepsilon_{FH,1}$, and $\varepsilon_{FH,2}$ is the Flory-Huggins parameters between solvent and uncharged monomers.

The isothermal adsorption free energy per unit of volume owing to the mixing of surfactant ions and monovalent polyelectrolyte counterions on ionic monomers is given by $G_{ads}$, see the third line of **Equation (1)**. The fraction of ionic monomers occupied preferentially by the large surfactant ions is denoted by $\varphi$. The excess-adsorption strength of one surfactant ion compared with a monovalent counterion on the ionic monomers is denoted by $\varepsilon_1$. The exchange chemical potential of a surfactant ion on the polymer chains is denoted by $\mu$, which scales as $\mu \sim \ln(nc_x)$ if the surfactant concentration $(n+1)c_x$ in bulk is very small. Because there is a strong demixing tendency between surfactant and solvent molecules, we take care of this demixing effect on polyelectrolyte chains by the parameter $\chi_s$ which can be as large as 2.0 for alkyl trimethyl ammonium bromides **[50]**. We denote the volume ratio between the ionic head of a surfactant and a solvent (or a counterion) by $\lambda$, which is usually on the order of unity though it can be varied significantly in some specific experiments **[51, 52]**.

As sketched in **Figure 1b**, we consider the associative attraction, $G_{attr}$, between ionized monomers caused by temporary bridges due to the hydrophobic-aggregation effect of adsorbed surfactant ions on polymer chains, see the fourth line of **Equation (1)**. Here the bridge is a kind of *short-range* attractive interaction deriving from the formation of electrostatic dipoles between monovalent surfactant ions and monovalent ionic monomers, which leads us to a rather simple statistical construction of $G_{attr}$. We note that unbound surfactant molecules can also assemble hydrophobically with a bound surfactant molecule. Thus, the presence of a free surfactant molecule will inevitably frustrate the effective temporary cross-linking effect. Under a mean-field consideration, we assume that a bound surfactant molecule meets a second bound surfactant molecule with a probability $(\varphi^2 c)$. However, the effective bridge can be formed only if free surfactant molecules are not present, which is given by the probability $(1 - \rho)$, where $\rho$ is the volume fraction of the surfactant ion next to polyelectrolyte chains in the solvent phase. The coupling energy $(\varepsilon_2)$ strands for the hydrophobic aggregation between surfactant ions that are adsorbed on polymer chains, which can be viewed as a constant without restricting generality.

The preferential adsorption of surfactant ions on ionic monomers inevitably leads to the enrichment of surfactant ions around polyelectrolyte chains, which indicates that there is a relation of $\rho \gg nc_x$. To simplify our following analytical calculations, this fact prompts us to assume that the saturation of the bound surfactant on polyelectrolyte chains is largely govern by the number of the already adsorbed surfactant, in other words we use the approximation $\rho \to \varphi$ in the construction of $G_{attr}$. This approximation essentially takes advantage of the fact that the electrostatic correlation between ionized monomers is highly screened and only short-range dipole interactions remain because of preferential adsorption of surfactant ions on ionic monomers. This approach will avoid heavy calculations without compromising on the physical conclusions. Notice that a physical boundary condition is explicitly embedded in the approximation $\rho \to \varphi$, that is both $\varphi \to 0$ and $\varphi \to 1$ leading to the vanishing of $G_{attr} \to 0$. The approximation correctly reflects the fact that polyelectrolyte is miscible within aqueous solutions with low and high concentrations of oppositely-charged surfactants in reentrant condensation respectively.

The surfactant concentration to occur the reentrant condensation of polyelectrolyte is usually less than about 100 mmol/L **[9-12]**. This fact indicates that the free energy of non-associative pairwise-like electrostatic interaction due to the *long-range* correlation of all ions can be approximated by the classical "double screening theory" **[53, 54]** in the low-salt limit as $G_{DS}$ with its truncated form, see the fifth line of **Equation (1)**. The inverse Debye screening length $\kappa$ is given by $(\kappa a)^2 = 4\pi(l_B/a)(pc + (n+1)c_x)$ for polyelectrolytes in diluted surfactant solutions, where $l_B$ is the Bjerrum length of water. There is a strong charge compensation/neutralization effect **[3, 7]** when monovalent surfactant ions adsorb on monovalent ionic monomers because of forming electrostatic dipoles. This results in the contribution of the electrostatic repulsion between ionized monomers for $G_{DS}$ is insignificant compared with other terms in free energy because the charge–charge interaction between ionized monomers is highly screened (see a similar argument in a recent excellent monograph by Muthukumar **[13]**). Therefore, in the following discussion we will ignore it to simplify analytical calculations.

**Table 1.** List of symbols used in the theory.

| Symbols | Physical meaning of the symbol |
|---|---|
| $N$ | the number of monomers in a polyelectrolyte chain |
| $n$ | the surfactant chain size in unit of a polyelectrolyte monomer |
| $a$ | the size of a monomer along the direction of the polymer backbone, the size of a solvent molecule, the size of a surfactant counterion and the size of a polyelectrolyte counterion |
| $p$ | the fraction of charged monomers in a polyelectrolyte chain |
| $\Pi$ | the osmotic pressure in the bulk |
| $c$ | the volume fraction of monomers |
| $c_x$ | the volume fraction of counterion of surfactant |
| $k_B$ | the Boltzmann constant |
| $T$ | the thermodynamic temperature |
| $\varphi$ | the fraction of ionic monomers occupied preferentially by the surfactant ions |
| $\lambda \approx 1$ | the volume ratio between the ionic head of a surfactant ion and a solvent (or a counterion) |
| $\chi_s$ | the parameter considers the demixing effect between surfactant and water on polyelectrolyte chains |
| $\mu$ | the exchange chemical potential for a surfactant ion on the polymer chains |
| $nc_x \ll \rho \to \varphi$ | the volume fraction of the surfactant ion next to polyelectrolyte chains in the solvent phase |
| $l_B$ | the Bjerrum length of water |
| $\kappa$ | the inverse Debye screening length |
| $\varepsilon_1$ | the excess-adsorption strength of one surfactant ion with respect to the monovalent counterion on ionic monomers |
| $\varepsilon_2$ | the hydrophobic-aggregation strength among surfactant ions that are adsorbed on ionic monomers |
| $\varepsilon_{FH,1}$ | the Flory-Huggins parameter between solvent and charged monomers |
| $\varepsilon_{FH,2}$ | the Flory-Huggins parameter between solvent and uncharged monomers |
| $G_{sol}$ | the free energy owing to the mixing of polymer with solvent and added surfactant |
| $G_{ads}$ | the adsorption free energy owing to the mixing of surfactant ions and polyelectrolyte counterions on the polymer chains |
| $G_{attr}$ | the associative attraction between ionized monomers caused by temporary bridge due to the hydrophobic-aggregation effect of adsorbed surfactant ions on ionic monomers |
| $G_{DS}$ | the free energy of non-associative pairwise-like electrostatic interaction due to the *long-range* correlation of all ions in the low-salt limit |
| $G_{FH}$ | the energy of non-electrostatic excluded-volume interactions between solvent (water) and the charge neutral part of monomers |
| $G(\varphi, c, c_x)$ | the total Gibbs free energy per volume unit under the framework of the isothermal–isobaric ensemble (NPT ensemble**)** |

Then, the total Gibbs free energy per volume unit is simply considered as $G(\varphi, c, c_x)$ = $G_{sol} + G_{FH} + G_{ads} + G_{attr} + G_{DS}$. We list the symbols used in the theory in **Table 1** for the reader's convenience. Notice that in this study we deliberately neglect the possible hydrophobic adsorption between the surfactant tail and a hydrophobic polymer backbone as well as the related electrostatic repulsion effect among adsorbed surfactant ions. This approach will significantly simplify our analytical calculations and guide us to focusing on the well-accepted phase-transition mechanism **[12, 15-18]** as illustrated by **Figure 1b**, and will finally lead us to qualitative but rather simple answers for the two fundamental problems raised in the section of **Introduction**.

## 3. Results and Discussion

## 3.1 The minimum coupling energy for the hydrophobic-aggregation effect in phase transition

Following a similar computing approach as our previous work **[55]**, we can estimate a minimum coupling energy ($\varepsilon_2$) for the hydrophobic-aggregation effect that is necessary for a phase transition. Based on the construction of $G_{attr}$ in **Equation(1)**, we see that the maximum coupling is achieved at the point $\varphi = 2/3$. This mathematically casts our model to a canonical ensemble-like model for the case of very diluted solution of surfactant ($(n+1)c_x \to 0$), where both the chemical potential $\mu$ and the osmotic pressure $\prod$ approach their limiting values $\mu_0$ and $\prod_0$ respectively:

$$G - \Pi_0 = \frac{c}{N}\ln(c) + pc\ln(pc) + \left[1-(1+p)c\right]\ln\left[1-(1+p)c\right]$$
$$+ \left[\frac{2\ln 2}{3\lambda} - \left(\frac{2}{3\lambda}+\frac{1}{3}\right)\ln 3 - \frac{2}{3}\left(\mu_0 + \frac{\varepsilon_1}{\lambda}\right) + \frac{2}{9}\chi_s\right]pc$$
$$-\frac{4}{27}\varepsilon_2(pc)^2 - \sqrt{\pi}\left(\frac{l_B}{a}pc\right)^{\frac{3}{2}}$$
$$+\left[1-(1+p)c\right]\left[p\varepsilon_{FH,1}+(1-p)\varepsilon_{FH,2}\right]c$$
(2)

Here the canonical ensemble-like free energy is given by $G - \prod_0$.

To show a noticeable hydrophobic effect, the surfactant chain length has to be above a minimum length. This implies that there exists a minimum coupling energy for the hydrophobic-aggregation effect ($\varepsilon_2$) that is necessary for a phase transition. By **Equation(2)**, we can roughly estimate a minimum value for the coupling energy ($\varepsilon_2$). This can be realized by determining the boundary condition for the spinodal decomposition of polyelectrolyte solution, which is given by $d^2(G - \prod_0)/dc^2 = 0$,

$$0 = \frac{d^2(G-\Pi_0)}{dc^2} = -\frac{8}{27}\varepsilon_2 p^2 - 2\varepsilon_{FH,1}(1+p)p - 2\varepsilon_{FH,2}(1-p^2)$$
$$+\frac{1}{Nc}+\frac{p}{c}+\frac{(1+p)^2}{1-(1+p)c}-\frac{3\sqrt{\pi}}{4}\left(\frac{l_B}{a}p\right)^{\frac{3}{2}}c^{-\frac{1}{2}}$$
(3)

Here we define an overall effective interaction parameter $2\chi_0 \equiv 8\varepsilon_2 p^2/27 + 2\varepsilon_{FH,1}(1+p)p$

+ $2\varepsilon_{FH,2}(1-p^2)$. The critical or minimum value of it to allow phase transition is given by $d(2\chi_0)/dc = 0$,

$$0 = \frac{d(2\chi_0)}{dc} = -\frac{1}{Nc^2} - \frac{p}{c^2} + \frac{(1+p)^3}{\left[1-(1+p)c\right]^2} + \frac{3\sqrt{\pi}}{8}\left(\frac{l_B}{a}\frac{p}{c}\right)^{\frac{3}{2}} \quad (4)$$

There exists no exact explicit analytical solution for $c$ in **Equation(4)**, but a good approximation solution can be obtained by ignoring the fourth term since it is much smaller than the sum of other terms within experimental values of the parameter $l_B/a$. For the critical point of the polyelectrolyte collapse, we obtain the approximation for the critical volume fraction of monomers $c$ at:

$$\frac{1}{c} \simeq (1+p) + \frac{(1+p)^{\frac{3}{2}}}{\left(\frac{1}{N}+p\right)^{\frac{1}{2}}} \quad (5)$$

The exact solution of **Equation(4)** is recovered by **Equation(5)** when the parameter $p$ approaches zero. One can improve the analytical solution of **Equation(4)** by using the method of fixed-point iteration **[56, 57]** with the initial value defined by **Equation(5)**. However, the approximate solution of **Equation(5)** is sufficient to deduce the key features of the overall effective interaction parameter $\chi_0$ without compromising on the physical conclusions.

By insertion of **Equation(5)** into **Equation(3)**, we get an estimation of the critical or minimum value of $\chi_0$,

$$2\chi_{0,min} \equiv \frac{8}{27}\varepsilon_2 p^2 + 2\varepsilon_{FH,1}(1+p)p + 2\varepsilon_{FH,2}(1-p^2)$$

$$\simeq \left[\left(\frac{1}{N}+p\right)^{\frac{1}{2}}(1+p)^{\frac{1}{2}} + (1+p)\right]^2 - \frac{3\sqrt{\pi}}{4}\left(\frac{l_B}{a}p\right)^{\frac{3}{2}}\left[(1+p) + \frac{(1+p)^{\frac{3}{2}}}{\left(\frac{1}{N}+p\right)^{\frac{1}{2}}}\right]^{\frac{1}{2}} \quad (6)$$

**Equation(6)** shows that a shorter polymer chain ($N$) prescribes a higher minimum coupling energy ($\varepsilon_2$) that necessary for a phase transition. This correctly reflects the obvious but nontrivial fact that a shorter polymer chain requires a higher enthalpic attraction among polymer chains to overcome/compensate a larger translational entropy of polymer chains in phase transition. By setting $p = 0$ in **Equation(6)**, we recover the boundary condition for the spinodal decomposition of an uncharged polymer solution, i.e., $\varepsilon_{FH,2} = \frac{1}{2}\left(1 + \frac{1}{\sqrt{N}}\right)^2$. Then we get the minimum of $\varepsilon_2$ for the case of a very long polyelectrolyte chain ($N \to \infty$) by **Equation(6)**,

$$\left(\varepsilon_2\right)_{\min} \simeq \frac{27}{8}\left[\left(\frac{1+p}{p}\right)^{\frac{1}{2}} + \frac{1+p}{p}\right]^2 - \frac{27\left[\varepsilon_{FH,1}(1+p)p + \varepsilon_{FH,2}(1-p^2)\right]}{4p^2} \\ - \frac{81\sqrt{\pi}}{32}\left(\frac{l_B}{a}\right)^{\frac{3}{2}}\left[\frac{1+p}{p} + \left(\frac{1+p}{p}\right)^{\frac{3}{2}}\right]^{\frac{1}{2}} \tag{7}$$

The strength of the hydrophobic-aggregation effect obviously depends on the surfactant chain length. It is also well-known that the strength of hydrophobic attraction is about 1.5 $k_B T$ between two methyl/methylene groups **[58]**. Therefore, there is a minimum surfactant chain length ($n_{min}$) **[44-47]** to induce phase separation of polyelectrolyte, which is numerically estimated by $n_{min} \approx 2(\varepsilon_2)_{\min}/3$ according to **Equation(6)** and **Equation(7)**. As an example of *hydrophilic* polyelectrolyte, we choose model parameters $p \approx 0.9$, $l_B/a \approx 2$, $N = 6000$ (or $N = 30$) and $\varepsilon_{FH,1} = \varepsilon_{FH,2} \approx 0.41$ for ionized poly(acrylic acid) **[59, 60]**, then **Equation(6)** predicts that the minimum surfactant chain length to induce phase separation is about $n_{min} \approx 7$ for the mixtures of ionized poly(acrylic acid) and alkyl trimethylammonium bromides. It is noticeable that this prediction is quite close to its experimental value ($n_{min} \approx 8$) **[45]** for the chosen set of model parameters.

A feature of the constructions of **Equation(6)** and **Equation(7)** is that the minimum coupling energy $(\varepsilon_2)_{\min}$ for the hydrophobic-aggregation effect estimated by **Equation(6)** and **Equation(7)**, is not related to the model parameters $\varepsilon_1$, $\lambda$ and $\chi_s$. This is an expected result, because one cannot expect a phase transition merely according to the simple exchange effect of surfactant ions on polymer chains based on a formal analogy with the well understood 1D-Ising model **[61, 62]**. Here, if we solely take into account pairwise-like interactions for preferential adsorption between surfactant ions and ionic monomers, linear polymers are 1D substrates for ions.

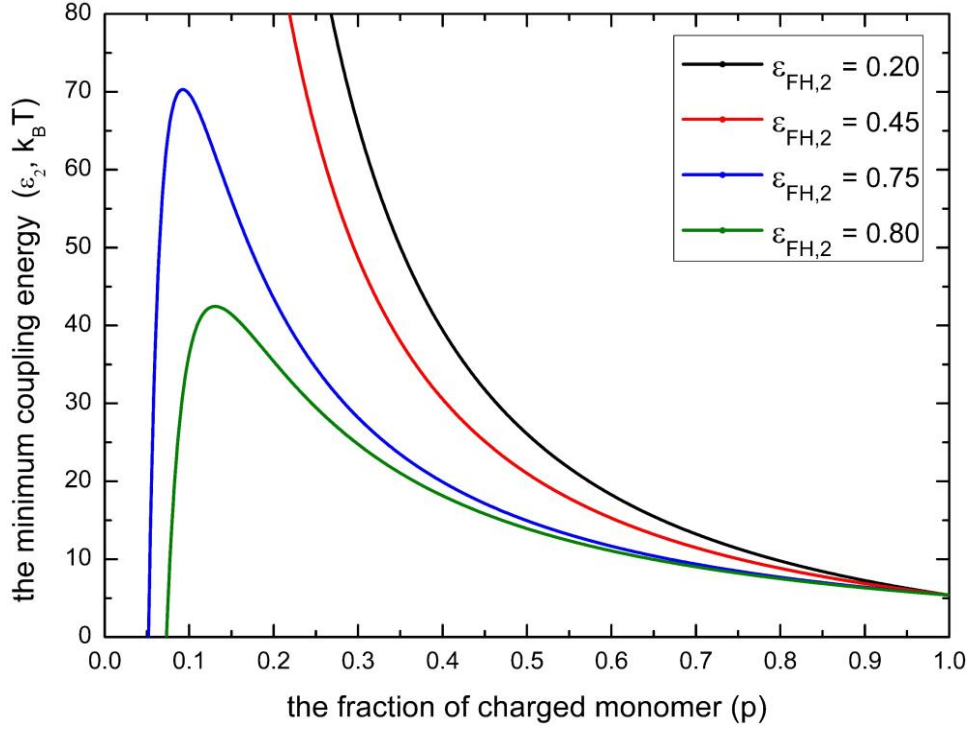

**Figure 2.** According to **Equation(7)**, the minimum coupling energy ($\varepsilon_2$) with respect to the fraction of charged monomer ($p$) for typical values of the parameter $\varepsilon_{FH,2}$ with $l_B/a = 2$, $\varepsilon_{FH,1} = 0.45$ and $N \to \infty$.

Another feature of the constructions of **Equation(6)** and **Equation(7)** is that: As shown in **Figure 2**, no matter the values of $l_B/a \geq 0$ and $\varepsilon_{FH,1} \geq 0$, there is a unique local maximum of $(\varepsilon_2)_{min}$ when $\varepsilon_{FH,2}$ is larger than $\frac{1}{2}\left(1 + \frac{1}{\sqrt{N}}\right)^2$, but no local maximum of $(\varepsilon_2)_{min}$ exists when $0 \leq \varepsilon_{FH,2} \leq \frac{1}{2}\left(1 + \frac{1}{\sqrt{N}}\right)^2$. This feature is coincidental with the well-known $\Theta$-condition for uncharged polymers. Thus, we can see that the minimum coupling energy $(\varepsilon_2)_{min}$ for the hydrophobic-aggregation effect to lead to a phase transition is related to the solvent quality (parameter $\varepsilon_{FH,2}$) for the uncharged part of polyelectrolyte. We note that this analytical result reflects the experimental fact **[12, 29]** that the hydrophobicity of uncharged monomers of polyelectrolytes has a noticeable effect on the necessary surfactant concentration to trigger the reentry transition.

### 3.2 The non-monotonic effective Flory–Huggins $\chi$ parameter

In order to find the equilibrium state of the polyelectrolyte phase with respect to the bulk solvent phase, we follow a similar computing approach as our previous work **[55]**. Firstly we minimize the free energy $G(\varphi, c, c_x)$ with respect to the adsorption fraction of surfactant ion $\varphi$. Then, the replacement of the solution $\varphi(\mu, c, c_x)$ into the free energy leads to an effective free energy for the polyelectrolyte, where the effect of surfactant ions is mapped onto an effective monomer–monomer interaction which will depend on the concentration of surfactant ions. Finally, the minimization of the free energy per monomer, i.e., $G(\varphi, c, c_x)/c$ instead of $G(\varphi, c, c_x)$, with respect to the monomer concentration ($c$) leads to the equilibrium solution of the model.

Because we are interested in the case of a very diluted solution of surfactant (($n$+1)$c_x$ → 0) in this work, we will ignore the influence of surfactant concentration in the bulk solution in the construction of $G_{sol}$, $G_{FH}$ and $G_{DS}$. This assumption allows us to focus on the adsorption and attraction effects of surfactant near/on polymer chains, and ignore their own non-essential mixing effects if surfactant molecules are far away from polymer chains. This approach will avoid heavy calculations without losing generality to get key physical conclusions.

**Equation(1)** implies that the maximum contraction of polymer chains is reached at the point $\varphi$ = 2/3, where the hydrophobic-aggregation effect reaches its maximum, which corresponds to the phase transition from soluble to collapsed states (including both collapse and reentry transitions). This peculiar feature of the model leads us to introducing a perturbation ($\delta$) from the two-thirds occupation of the chain by the surfactant ion according to

$$\varphi = \frac{2}{3}(1-\delta) \tag{8}$$

This perturbation approach follows closely with our previous work **[63]**. With the Taylor expansion of $\delta$-containing terms in the logarithm function up to the accuracy of square terms ($\delta^2$) under the constraint of $|\delta^3| \ll 1$ and ignoring constant terms, we obtain

$$\frac{G(\delta,c)}{c} = \left(\frac{G_{sol}+G_{DS}+G_{FH}}{c}\right) + \frac{2}{3}\left(\mu + \frac{\varepsilon_1 + \ln 3 - \ln 2}{\lambda} - \ln 3\right)p\delta \\ + \frac{2}{3}\left(1-\frac{1}{\lambda}\right)p\delta + \frac{1}{3}\left(2+\frac{1}{\lambda}\right)p\delta^2 + \frac{2}{9}\chi_s\left(\delta - 2\delta^2\right)p - \frac{4}{27}\varepsilon_2\left(1-3\delta^2\right)p^2 c \tag{9}$$

with

$$\frac{G_{sol}+G_{DS}+G_{FH}}{c} = \frac{\ln(c)}{N}+p\ln(pc)+\left[\frac{1}{c}-(1+p)\right]\ln\left[1-(1+p)c\right]$$
$$+\left[1-(1+p)c\right]\left[p\varepsilon_{FH,1}+(1-p)\varepsilon_{FH,2}\right]-\sqrt{\pi}\left(\frac{l_B}{a}p\right)^{\frac{3}{2}}\sqrt{c}+\frac{\Pi}{c} \quad (10)$$

The numerical term "ln2" in **Equation(9)** due to the symmetry of adsorption and desorption states of ionic monomers is broken. This can be seen from the transformation of $G_{ads}$ in **Equation(1)** by the manipulation of **Equation(8)**.

Minimizing the free energy in **Equation(9)** with respect to $\delta$ yields

$$\delta = -\frac{3\left(\mu+\frac{\varepsilon_1+\ln 3-\ln 2}{\lambda}-\ln 3\right)+\chi_s+3\left(1-\frac{1}{\lambda}\right)}{4\varepsilon_2 pc+3\left(2+\frac{1}{\lambda}\right)-4\chi_s}. \quad (11)$$

Resubstitute **Equation(11)** into **Equation(9)**, we obtain

$$\frac{G(c)}{c} = -\frac{4}{27}\varepsilon_2 p^2 c - \frac{p\left[\left(\mu+\frac{\varepsilon_1+\ln 3-\ln 2}{\lambda}-\ln 3\right)+\frac{\chi_s}{3}+\left(1-\frac{1}{\lambda}\right)\right]^2}{4\varepsilon_2 pc+3\left(2+\frac{1}{\lambda}\right)-4\chi_s} + \left(\frac{G_{sol}+G_{DS}+G_{FH}}{c}\right)$$

$$= p\varepsilon_{FH,1}+(1-p)\varepsilon_{FH,2}+p\ln(p)-\frac{p\left[\left(\mu+\frac{\varepsilon_1+\ln 3-\ln 2}{\lambda}-\ln 3\right)+\frac{\chi_s}{3}+\left(1-\frac{1}{\lambda}\right)\right]^2}{3\left(2+\frac{1}{\lambda}\right)-4\chi_s} \quad (12)$$

$$-\chi_{eff}c+g_{sol}$$

with the effective Flory–Huggins parameter $\chi_{eff}$:

$$\chi_{eff} = \varepsilon_{FH,1}(1+p)p+\varepsilon_{FH,2}(1-p^2)+\frac{4}{27}\varepsilon_2 p^2$$
$$-(4\varepsilon_2 p^2)\frac{\left[\left(\mu+\frac{\varepsilon_1+\ln 3-\ln 2}{\lambda}-\ln 3\right)+\frac{\chi_s}{3}+\left(1-\frac{1}{\lambda}\right)\right]^2}{\left[4\varepsilon_2 pc+3\left(2+\frac{1}{\lambda}\right)-4\chi_s\right]\left[3\left(2+\frac{1}{\lambda}\right)-4\chi_s\right]}+\sqrt{\pi}\left(\frac{l_B}{a}p\right)^{\frac{3}{2}}\frac{1}{\sqrt{c}} \quad (13)$$

and the entropic term of free energy per monomer $g_{sol}$:

$$g_{sol} = \frac{\ln(c)}{N}+p\ln(c)+\left[\frac{1}{c}-(1+p)\right]\ln\left[1-(1+p)c\right]+\frac{\Pi}{c} \quad (14)$$

The above constructed $\chi$-function is a function of the square of chemical potential $\mu$, which indicates that a $\chi_{eff}$ corresponds to two values of $\mu$ and thus essentially captures the reentrant signature of polyelectrolyte condensation at lower and higher surfactant concentrations. According to **Equation(13)** we know that the reentrant signature of polyelectrolyte condensation is controlled by the hydrophobic-aggregation effect, since only this effect is non-monotonic with respect to the surfactant concentration.

### 3.3 The general spinodal phase diagrams

In order to analytically discuss the phase transition, the pressure isotherm $\Pi(c, \mu; N, l_B/a, p, \varepsilon_1, \varepsilon_2, \varepsilon_{FH,1}, \varepsilon_{FH,2})$ can be calculated from **Equation(12)** by $\partial(G/c)/\partial c = 0$, which leads to

$$\Pi = 4\varepsilon_2 p^2 c^2 \left\{ \frac{\left[\left(\mu + \frac{\varepsilon_1 + \ln 3 - \ln 2}{\lambda} - \ln 3\right) + \frac{\chi_s}{3} + \left(1 - \frac{1}{\lambda}\right)\right]^2}{\left[4\varepsilon_2 pc + 3\left(2 + \frac{1}{\lambda}\right) - 4\chi_s\right]^2} - \frac{1}{27} \right\} + \left(\frac{1}{N} - 1\right)c$$
$$- \left[p(1+p)\varepsilon_{FH,1} + (1-p^2)\varepsilon_{FH,2}\right]c^2 - \frac{\sqrt{\pi}}{2}\left(\frac{l_B}{a}pc\right)^{\frac{3}{2}} - \ln\left[1 - (1+p)c\right] \quad (15)$$

By setting $p = 0$ in **Equation(15)** for uncharged polymer solution, with $\partial\Pi/\partial c = 0$ and the constraint of $0 < c < 1$, as it should be, we recover the boundary condition for the spinodal decomposition of uncharged polymer solution, i.e., $\varepsilon_{FH,2} > \frac{1}{2}\left(1 + \frac{1}{\sqrt{N}}\right)^2$.

For the general case of a discontinuous phase transition, the osmotic pressure must display an unstable region of negative compressibility given by $\partial\Pi/\partial c < 0$. In **Figure 3** we display an example which shows the pressure isotherms for a discontinuous condensation transition given by **Equation(15)**. The coexistence region of binodal is defined by the Maxwell construction as indicated by the equal area criterion (the area of the envelope "$ABC$" is equal to that of the envelope "$CDE$": $S_1 = S_2$) with the horizontal isobaric line in the figure (the black dashed line "$ACE$"). This cannot be obtained analytically in an exact way and will not be considered in details in the following discussions if not otherwise noted specifically. However, we are able to calculate the spinodal analytically at which the solution starts to become unstable and the existence of which is the necessary condition for a discontinuous transition scenario. The spinodal of polyelectrolyte solution is given by $\partial\Pi/\partial c = 0$ and can be written in the following form:

$$U^2 = \frac{\left[4\varepsilon_2 pc + 3\left(2 + \frac{1}{\lambda}\right) - 4\chi_s\right]^3}{8p^2\varepsilon_2 c\left[3\left(2 + \frac{1}{\lambda}\right) - 4\chi_s\right]} \left\{ \frac{8\varepsilon_2 p^2}{27}c + 2\left[p(1+p)\varepsilon_{FH,1} + (1-p^2)\varepsilon_{FH,2}\right]c + \frac{3\sqrt{\pi}}{4}\left(\frac{l_B}{a}p\right)^{\frac{3}{2}}\sqrt{c} - \frac{(1+p)}{1-(1+p)c} + \left(1 - \frac{1}{N}\right) \right\} \quad (16)$$

Here, we denote the external adsorption field $U$ as:

$$U \equiv \left(\mu + \frac{\varepsilon_1 + \ln 3 - \ln 2}{\lambda} - \ln 3\right) + \frac{\chi_s}{3} + \left(1 - \frac{1}{\lambda}\right) \quad (17)$$

This defines the spinodal phase diagram in the "$\mu-c$" space with the eight parameters $p$, $N$, $l_B/a$, $\varepsilon_1$, $\varepsilon_2$, $\chi_s$, $\varepsilon_{FH,1}$ and $\varepsilon_{FH,2}$.

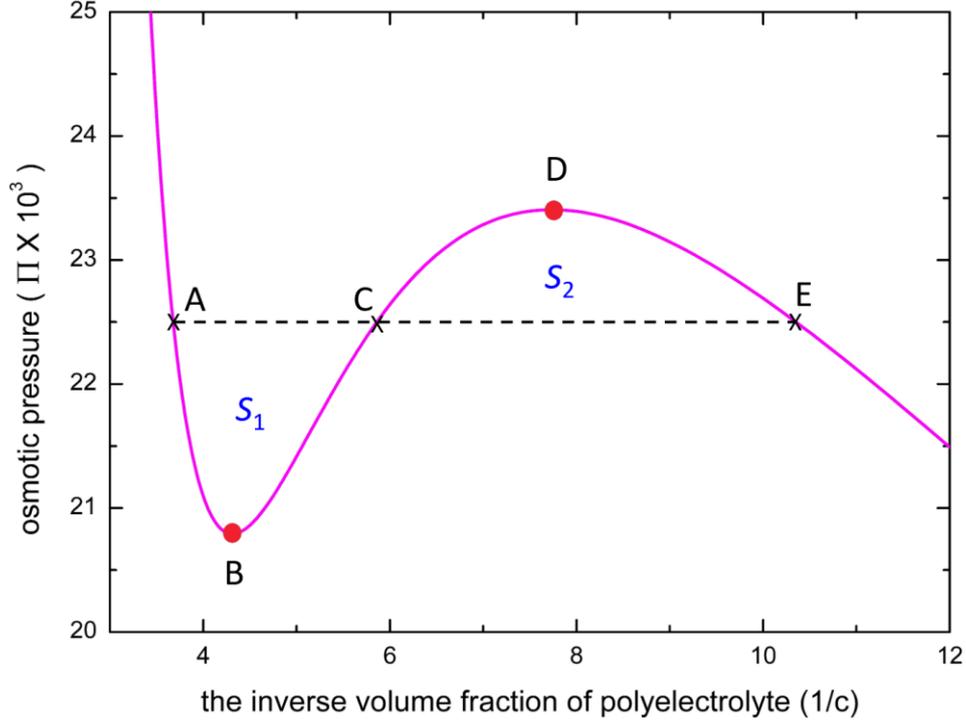

**Figure 3.** According to **Equation(15)**, osmotic pressure of polyelectrolyte solution is plotted as a function of inverse volume fraction of monomer for the parameters $p = 0.6$, $l_B/a = 2$, $\mu = -6$, $\varepsilon_1 = 6$, $\varepsilon_2 = 15$, $\varepsilon_{FH,1} = 0.45$, $\varepsilon_{FH,2} = 0.45$, $\lambda = 1$, $\chi_s = 1$ and $N \to \infty$. The coexistence pressure of binodal by the Maxwell construction is indicated by the horizontal dashed line in the figure (the black dashed line "*ACE*") with the equal area criterion (the area of the envelope "*ABC*" is equal to that of the envelope "*CDE*": $S_1 = S_2$), and the spinodal points are indicated by filled circles in the figure (symbols "*B*" and "*D*" in the figure).

In **Figure 4** we display the spinodal phase diagram in the "$\mu-c$" space given by **Equation(16)** for a typical case of polyelectrolyte with parameters $p = 0.6$, $l_B/a = 2$, $\varepsilon_1 = 6$, $\varepsilon_2 = 15$, $\varepsilon_{FH,1} = \varepsilon_{FH,2} = 0.4$, $\lambda = 1$, $\chi_s = 2$ and $N \to \infty$. The two solutions at the same value of $\mu$ correspond to the two extrema of the pressure isotherm, which correspond to the coexistence of a condensed polymer phase and a dissolved polymer phase. As an example in **Figure 4**, the blue dashed line "*ABC*" in the figure shows how to determine the coexistence concentrations of two polymer phases by the lever rule. The region of phase separation is closed topologically. For our theory, we obtain a symmetric collapse and reentry transition. The lower part of $\mu$ defines the collapse transition as indicated by the lower half of the "egg-shape" curve, while the higher part of $\mu$ defines the reentry transition.

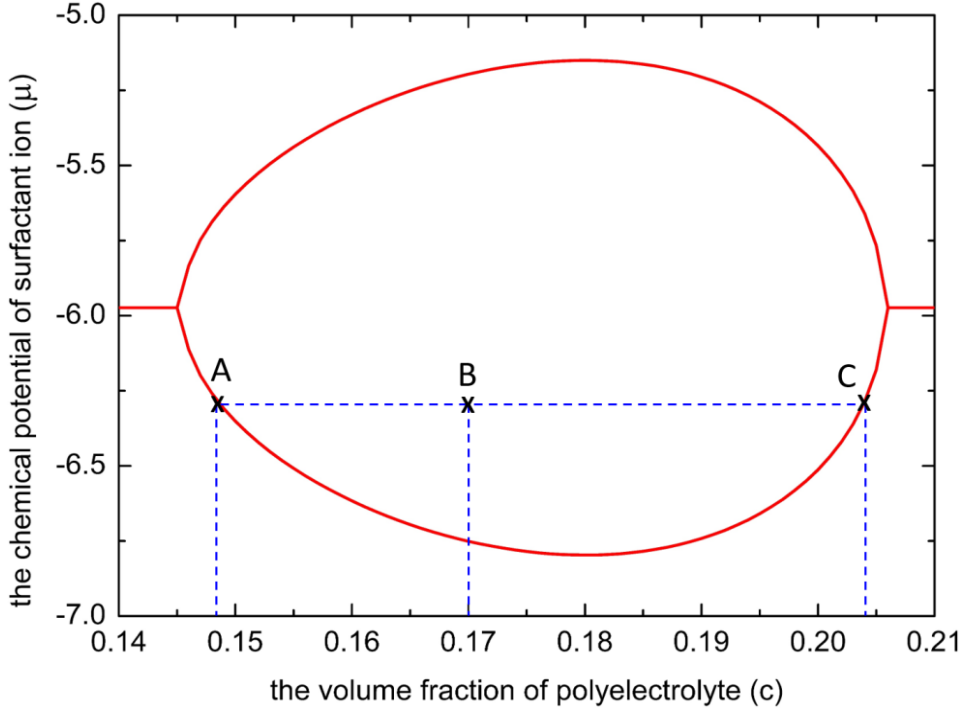

**Figure 4.** According to **Equation (16)**, a typical spinodal phase diagram in the "$\mu-c$" space for a polyelectrolyte in the dilute solution of an oppositely-charged surfactant for the case of $p = 0.6$, $l_B/a = 2$, $\varepsilon_1 = 6$, $\varepsilon_2 = 15$, $\varepsilon_{FH, 1} = \varepsilon_{FH, 2} = 0.4$, $\lambda = 1$, $\chi_s = 2$ and $N \to \infty$. The blue dashed line "*ABC*" in the figure shows an example how to determine the coexistence concentrations of two polymer phases by the lever rule.

### 3.4 Comparison to experiments

In this subsection, we compare analytical predictions of our theory with experimental results reported in literature.

In **Figure 5** we display the spinodal phase diagram according to **Equation(16)** for various polyelectrolyte chain lengths ($N$) under the condition of $p = 0.6$, $l_B/a = 2$, $\varepsilon_1 = 6$, $\varepsilon_2 = 18$, $\varepsilon_{FH, 1} = \varepsilon_{FH, 2} = 0.4$, $\lambda = 1$ and $\chi_s = 1$. As shown in **Figure 5**, phase transition can be influenced drastically by the chain length ($N$) of a polyelectrolyte. We see that an increase of polyelectrolyte chain length will promote the coexistence region of collapse transition to a lower surfactant concentration, and can shift the coexistence region of reentry transition to a higher surfactant concentration. However, this chain-length effect disappears after the polyelectrolyte chain length reaches a maximum length. In contrast to uncharged linear polymer **[13, 48, 49]**, by setting the external adsorption field $U$ = 0 in **Equation (16)**, our model shows that it is hard to realize a real dilute phase for the reentrant condensation of polyelectrolyte when the charge fraction is not sufficiently low. As indicated in **Figure 5,** this is particularly noticeable in the limiting case of infinite polyelectrolyte chain length. There remains no small monomer concentration ($c$) in dilute phase for small values of the external field $U$ when phase separation occurs, i.e., close to the optimally loaded state of the polyelectrolyte with surfactant ions where the effective Flory–Huggins parameter $\chi_{\text{eff}}$ reaches its maximum (see **Equation (13)**). Our theory also predicts that there is no phase transition if the polyelectrolyte chain length is too short. For example, the shortest polyelectrolyte chain length to induce phase separation is about $N$ = 10 for **Figure 5** with other parameters unchanged. Actually, this is a reflection that a shorter polyelectrolyte chain prescribes a higher minimum coupling energy for the hydrophobic-aggregation effect in phase transition, see **Equation (6)**. It is remarkable that all these analytical predictions of our theory were confirmed in detail by previous experimental studies **[46, 64-66]** on the influence of polyelectrolyte chain length in the phase separation of polyelectrolyte admixed with oppositely-charged surfactant.

In **Figure 6a**, we display the spinodal phase diagram according to **Equation(16)** for different strengths of hydrophobic-aggregation effect ($\varepsilon_2$) under the condition of $p$ = 0.6, $l_\text{B}/a$ = 2, $\varepsilon_1$ = 6, $\varepsilon_{\text{FH, 1}}$ = $\varepsilon_{\text{FH, 2}}$ = 0.4, $\lambda$ = 1, $\chi_\text{s}$ = 1 and $N \to \infty$. We see that an increase of the strength of hydrophobic-aggregation effect ($\varepsilon_2$), in other words an increase of surfactant chain length ($n$) will shift the coexistence region of collapse transition to a lower concentration of surfactant, but shift the coexistence region of reentry transition to a higher concentration of surfactant, which is in agreement with experimental results of highly-hydrophilic polyelectrolytes in aqueous solutions of oppositely-charged surfactants **[14, 44-47]**. However, when the surfactant chain length is longer than a maximum length, it is expected that the chain-length effect of surfactant will saturate due to the lack of an efficient stack of steric surfactant tails in their premicellar aggregation (see **Figure 1b** for a sketch of the premicellar aggregation). But in this situation, the demixing effect between surfactant and water ($\chi_\text{s}$) still plays a role on polyelectrolyte chains. In **Figure 6b** we display the spinodal phase diagram according to **Equation(16)** for different demixing effects between surfactant and water ($\chi_s$) under the condition of $p = 0.6$, $l_B/a = 2$, $\varepsilon_1 = 6$, $\varepsilon_2 = 15$, $\varepsilon_{FH, 1} = \varepsilon_{FH, 2} = 0.4$, $\lambda = 1$ and $N \to \infty$. We see that an increase of the demixing effect between surfactant and water ($\chi_s$) will shift the coexistence region of both collapse and reentry transitions to lower concentrations of surfactant. This analytical result indeed qualitatively explains the phase-behavior difference of a highly-hydrophilic polyelectrolyte in aqueous solutions of hexadecyl- and dodecyl- trimethylammonium bromides (both surfactants have a very long steric alkyl tail) **[12]**.

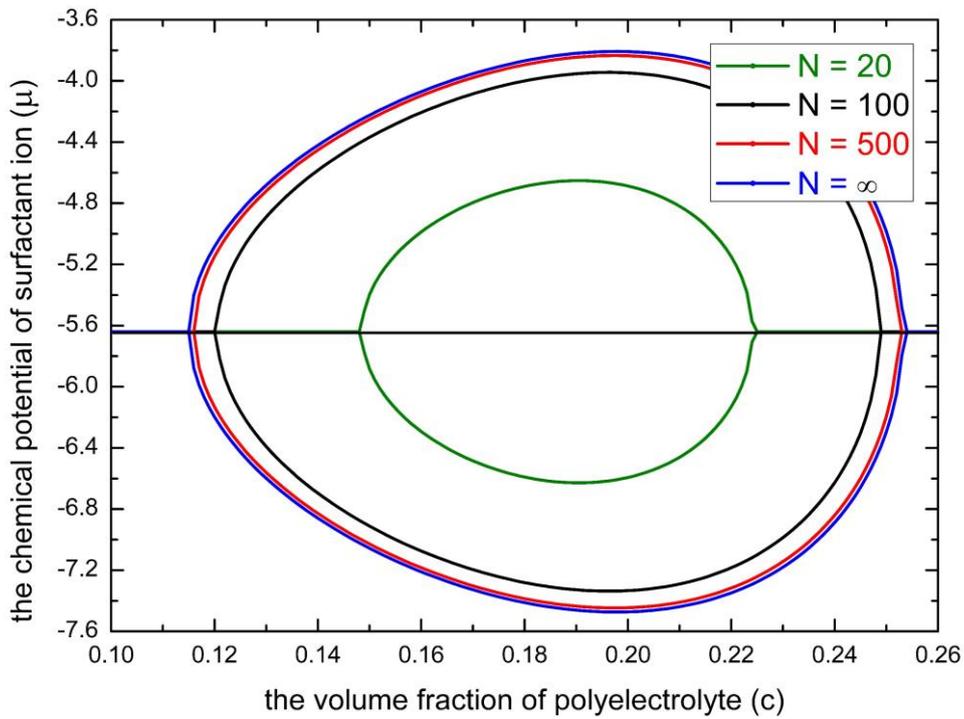

**Figure 5.** According to **Equation (16)**, spinodal phase diagrams of polyelectrolyte in the dilute solution of an oppositely-charged surfactant for different polyelectrolyte chain lengths ($N$) under the condition of $p = 0.6$, $l_B/a = 2$, $\varepsilon_1 = 6$, $\varepsilon_2 = 18$, $\varepsilon_{FH, 1} = \varepsilon_{FH, 2} = 0.4$, $\lambda = 1$ and $\chi_s = 1$.

Interestingly, by an increase in charge fraction ($p$) of a polyelectrolyte chain while keeping other model parameters unchanged, the corresponding spinodal phase diagrams according to our theory (similar to **Figure 6a**, also see **Equation(16)**) indicate that the coexistence region of collapse transition shifts to a lower concentration of surfactant, but the coexistence region of reentry transition shifts to a higher concentration of surfactant. It is noticeable that this analytical result was corroborated in detail by experimental studies of a recent dissertation **[67]** on the phase behaviors of poly(acrylic acid) and sodium polyacrylate in aqueous solutions of hexadecyl trimethylammonium bromide.

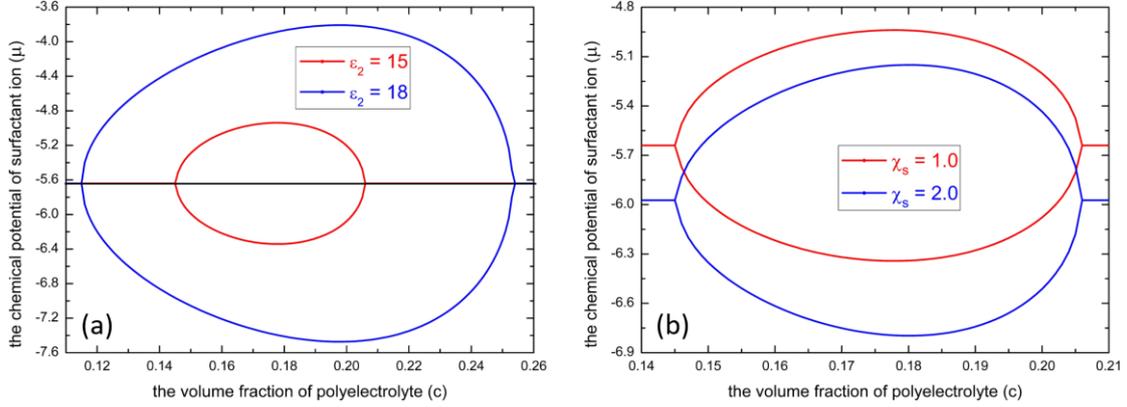

**Figure 6. (a)** According to **Equation (16)**, spinodal phase diagrams of polyelectrolyte in the dilute solution of an oppositely-charged surfactant for different strengths of hydrophobic-aggregation effect ($\varepsilon_2$) under the condition of $p = 0.6$, $l_B/a = 2$, $\varepsilon_1 = 6$, $\varepsilon_{FH,1} = \varepsilon_{FH,2} = 0.4$, $\lambda = 1$, $\chi_s = 1$ and $N \to \infty$. **(b)** According to **Equation (16)**, spinodal phase diagrams of polyelectrolyte in the dilute solution of an oppositely-charged surfactant for different demixing effects between surfactant and water ($\chi_s$) under the condition of $p = 0.6$, $l_B/a = 2$, $\varepsilon_1 = 6$, $\varepsilon_2 = 15$, $\varepsilon_{FH,1} = \varepsilon_{FH,2} = 0.4$, $\lambda = 1$ and $N \to \infty$.

We note that in the charge compensation/neutralization of admixing polyelectrolyte and an oppositely charged surfactant, it can form a strong electrostatic dipole between a monovalent ionic monomer and a monovalent surfactant ion because of a noticeable reduction of dielectric constant around polyelectrolyte chains **[13, 54]**. This indicates that there is a significant shift of $\varepsilon_1$ on the left-hand side of the spinodal construction, i.e., **Equation (16)**, which is often not small for polyelectrolyte (on the order of about 5 $k_B T$ for the strength of an ionic bond **[68]** in water at low salt concentration and in the order of about 10 $k_B T$ for the strength of electric dipole interaction **[69]**). This fact implies that **Equation (16)** melts down when the value of $\mu$ is far away from zero, which is approximately given by the condition that the external adsorption field $U$ is close to zero:

$$0 \approx U \equiv \left( \mu_0 + \frac{\varepsilon_1 + \ln 3 - \ln 2}{\lambda} - \ln 3 \right) + \frac{\chi_s}{3} + \left( 1 - \frac{1}{\lambda} \right) \quad (18)$$

By **Equation (18)**, we can estimate the critical aggregation concentration (CAC) of surfactant in the surfactant-polyelectrolyte binding isotherm for occurring a reentrant condensation, which is given by

$$\ln(\text{CAC}) \sim \mu_0 \approx -\frac{\varepsilon_1}{\lambda} - \frac{\chi_s}{3} + \frac{1+\ln 2 - \ln 3}{\lambda} + \ln 3 - 1 \tag{19}$$

We can read some interesting physics from **Equation (19)**. First, a larger size of the ionic head of a surfactant ion ($\lambda$) can promote the CAC to a higher surfactant concentration. It is remarkable that this analytical prediction was confirmed by previous experimental studies **[43, 51, 52]** on the reentrant condensation of some polyelectrolytes. Second, if the values of $\varepsilon_1$ and $\chi_s$ are large enough, for example, a rather moderate choice of parameters $\varepsilon_1 = 6$, $\chi_s = 1$ and $\lambda \approx 1$ leads to that the surfactant concentration for a polyelectrolyte reentrant condensation to occur is on the order of about $(n+1)c_x \sim \text{CAC} \approx 1.0$ mM/L. According to this simple evaluation, we know that the surfactant concentrations for polyelectrolyte reentrant condensation can be much lower than its bulk critical micelle concentration (CMC). This is the case for both the collapse and reentry branches of the reentrant condensation as shown from **Figure 4** to **Figure 6**. Third, because of $\lambda \approx 1$ and $\varepsilon_1 \gg \chi_s / 3$ for most common cases, by **Equation (19)** we see that the primary factor to induce both collapse and reentry transitions of polyelectrolyte at low surfactant concentrations, is due to the strong electrostatic adsorption between ionic monomers and surfactant ions ($\varepsilon_1$). Notice that these predictions concur with all experimental observations reported in the literature **[1-8]** on the reentrant condensation of polyelectrolyte in dilute aqueous solutions of oppositely-charged surfactants.

## 4. Conclusions

In summary, in this work we focus on the adsorption and attraction effects of surfactant near/on polymer chains, and ignore their own non-essential mixing effects if surfactant molecules are far away from polymer chains. This approach allows us to construct a simple mean-field theory and solve it analytically, and finally rationalize the essential features (such as the "egg shape" of spinodal phase diagrams) of the reentrant condensation of a polyelectrolyte induced by diluted oppositely-charged surfactants. By theoretical analysis, we found that a strong electrostatic adsorption between the ionic monomers and surfactant ions is critical to understand the peculiar phenomenon that both the collapse and reentry transitions of polyelectrolytes can occur when the concentration of surfactant is lower than its bulk critical micelle concentration (CMC). The analytical solution of the theory indicates that a minimum coupling energy for the nonlinear hydrophobic-aggregation effect of adsorbed surfactant is essential for phase transition to occur, which explains why polyelectrolytes show phase transition only if the surfactant chain length is longer than a minimum length [44-47].

Remarkably, the nonlinear attraction effect among ionic monomers deriving from the electrostatic adsorption of surfactant [12, 15-18] (as illustrated by **Figure 1b**), can be leveraged to explain/analyze some co-condensation behaviors of proteins with DNAs [24, 25]. For example, the phase separation of DNA induced by protein binding is dependent on the DNA length [19, 25], this can definitely be explained by our theory on the effect of polyelectrolyte chain length (see **Figure 5**). Here the protein behaves like an ionic surfactant and DNA is a bio-polyelectrolyte. It is thus natural to envisage that our theoretical results will shed light on a deep understanding of liquid-liquid phase separation in biological systems where ionic surfactant-like proteins/peptides (such as FUS and HP1 proteins) bound to bio-polyelectrolytes such as RNAs and DNAs play an important role, which remains under debate [19-27].

Nevertheless, we ought to point out that in the current theoretical formalism we have neglected the interaction between the polyelectrolyte backbone and the hydrophobic tail of surfactant. This implies that our simplified formalism is primarily confined to a hydrophilic polyelectrolyte backbone, i.e., in the case of $\varepsilon_{\text{FH},1} < \frac{1}{2}\left(1+\frac{1}{\sqrt{N}}\right)^2$ and $\varepsilon_{\text{FH},2} < \frac{1}{2}\left(1+\frac{1}{\sqrt{N}}\right)^2$. Whereas for polyelectrolyte backbones with strong hydrophobic properties, our theoretical approach can be flexibly refined and extended to take care of additional hydrophobic adsorption between the surfactant tail and polymer backbone as well as related electrostatic repulsion effect among adsorbed surfactant ions. These additional effects can affect and decease considerably the surfactant concentration that is necessary to occur the reentry transition as pointed out by previous experiments [12, 29]. However, a detailed investigation of this aspect in an analytical way is worthy of future consideration and lies beyond the scope of the present study.

## Author Information

**Corresponding Author**
Huaisong Yong, h.yong@utwente.nl (H. Y.); yonghuaisong@gmail.com (H. Y.)

**Notes**
The author declares no competing financial interest.

# Acknowledgments

The author acknowledges the partial financial support for this research by the Deutsche Forschungsgemeinschaft (DFG) under the project number 422913191, the partial financial support for this research by the "Tianfu Emei" Scholar Foundation of Sichuan Province (No.: 2326), and the visiting project for this research licensed by the Federal Office for Migration and Refugees of Germany (No.: BAMF 5-00002915). The author thanks Dr. Holger Merlitz at Leibniz-Institut für Polymerforschung Dresden, Dr. Qian Huang at Sichuan University and Dr. Jingguo Li at the University of Science and Technology of China for their insightful comments on the manuscript.

# Additional Note

The expression "dilute solution of a surfactant" in this research means that the concentration of an ionic surfactant is lower than its bulk critical micelle concentration (CMC), usually less than about 100 mmol/L [9-12]. This indicates that the Debye–Hückel theory remains qualitatively valid for the ionic surfactant, which requires that the ionic strength of the added small salt (such as the ionic surfactant) is low, usually on the order of 200 mmol/L[70].

# References

[1] A. Puente-Santamaría, F. Ortega, A. Maestro, R.G. Rubio, E. Guzmán, Non-equilibrium states in polyelectrolyte-surfactant systems at fluid interfaces: A critical review, Current Opinion in Colloid & Interface Science, 71 (2024) 101804.

[2] E. Guzmán, A. Maestro, F. Ortega, R.G. Rubio, Association of oppositely charged polyelectrolyte and surfactant in solution: equilibrium and nonequilibrium features, Journal of Physics: Condensed Matter, 35 (2023) 323001.

[3] M. Gradzielski, Polyelectrolyte–Surfactant Complexes As a Formulation Tool for Drug Delivery, Langmuir, 38 (2022) 13330-13343.

[4] S. Llamas, E. Guzmán, F. Ortega, N. Baghdadli, C. Cazeneuve, R.G. Rubio, G.S. Luengo, Adsorption of polyelectrolytes and polyelectrolytes-surfactant mixtures at surfaces: a physico-chemical approach to a cosmetic challenge, Advances in Colloid and Interface Science, 222 (2015) 461-487.

[5] L. Piculell, Understanding and Exploiting the Phase Behavior of Mixtures of Oppositely Charged Polymers and Surfactants in Water, Langmuir, 29 (2013) 10313-10329.

[6] L. Chiappisi, I. Hoffmann, M. Gradzielski, Complexes of oppositely charged polyelectrolytes and surfactants – recent developments in the field of biologically derived polyelectrolytes, Soft Matter, 9 (2013) 3896.

[7] C. La Mesa, Polymer–surfactant and protein–surfactant interactions, Journal of Colloid and Interface Science, 286 (2005) 148-157.

[8] A.F. Thünemann, Polyelectrolyte–surfactant complexes (synthesis, structure and materials aspects), Progress in Polymer Science, 27 (2002) 1473-1572.

[9] J. Carlstedt, D. Lundberg, R.S. Dias, B. Lindman, Condensation and Decondensation of DNA by Cationic Surfactant, Spermine, or Cationic Surfactant–Cyclodextrin Mixtures: Macroscopic Phase Behavior, Aggregate Properties, and Dissolution Mechanisms, Langmuir, 28 (2012) 7976-7989.

[10] D. Li, M.S. Kelkar, N.J. Wagner, Phase Behavior and Molecular Thermodynamics of Coacervation in Oppositely Charged Polyelectrolyte/Surfactant Systems: A Cationic Polymer JR 400 and Anionic Surfactant SDS Mixture, Langmuir, 28 (2012) 10348-10362.

[11] K. Bodnár, K. Szarka, M. Nagy, R. Mészáros, Effect of the Charge Regulation Behavior of Polyelectrolytes on Their Nonequilibrium Complexation with Oppositely Charged Surfactants, The Journal of Physical Chemistry B, 120 (2016) 12720-12729.

[12] K. Bodnár, E. Fegyver, M. Nagy, R. Mészáros, Impact of Polyelectrolyte Chemistry on the Thermodynamic Stability of Oppositely Charged Macromolecule/Surfactant Mixtures, Langmuir, 32 (2016) 1259-1268.

[13] M. Muthukumar, Physics of Charged Macromolecules: Synthetic and Biological Systems, Cambridge University Press, Cambridge, United Kingdom, 2023.

[14] N. Jain, S. Trabelsi, S. Guillot, D. McLoughlin, D. Langevin, P. Letellier, M. Turmine, Critical Aggregation Concentration in Mixed Solutions of Anionic Polyelectrolytes and Cationic Surfactants, Langmuir, 20 (2004) 8496-8503.

[15] P. Zheng, D. Cai, Z. Zhang, Y. Yang, T. Yin, W. Shen, Interactions between Sodium Polyacrylate and Mixed Micelles of Dodecyltrimethylammonium Bromide and Sodium Bis(2-ethylhexyl) Sulfosuccinate, Macromolecules, 46 (2012) 247-256.

[16] C. Wang, K.C. Tam, Interaction between Polyelectrolyte and Oppositely Charged Surfactant: Effect of Charge Density, The Journal of Physical Chemistry B, 108 (2004) 8976–8982.

[17] C. Wang, K.C. Tam, New Insights on the Interaction Mechanism within Oppositely Charged Polymer/Surfactant Systems, Langmuir, 18 (2002) 6484-6490.

[18] S. dos Santos, C. Gustavsson, C. Gudmundsson, P. Linse, L. Piculell, When Do Water-Insoluble Polyion−Surfactant Ion Complex Salts "Redissolve" by Added Excess Surfactant?, Langmuir, 27 (2010) 592-603.

[19] J. Park, J.-J. Kim, J.-K. Ryu, Mechanism of phase condensation for chromosome architecture and function, Experimental & Molecular Medicine, 56 (2024) 809-819.

[20] L. Nordenskiöld, X. Shi, N. Korolev, L. Zhao, Z. Zhai, B. Lindman, Liquid-liquid phase separation (LLPS) in DNA and chromatin systems from the perspective of colloid physical chemistry, Advances in Colloid and Interface Science, 326 (2024) 103133.

[21] S.P. Le, J. Krishna, P. Gupta, R. Dutta, S. Li, J. Chen, S. Thayumanavan, Polymers for Disrupting Protein–Protein Interactions: Where Are We and Where Should We Be?, Biomacromolecules, 25 (2024) 6229–6249.

[22] N. Akilli, T. Cheutin, G. Cavalli, Phase separation and inheritance of repressive chromatin domains, Current Opinion in Genetics & Development, 86 (2024) 102201.

[23] P.L. Toledo, A.R. Gianotti, D.S. Vazquez, M.R. Ermácora, Protein nanocondensates: the next frontier, Biophysical Reviews, 15 (2023) 515-530.

[24] R. Renger, J.A. Morin, R. Lemaitre, M. Ruer-Gruss, F. Jülicher, A. Hermann, S.W. Grill, Co-condensation of proteins with single- and double-stranded DNA, Proceedings of the National Academy of Sciences, 119 (2022).

[25] J.-K. Ryu, C. Bouchoux, H.W. Liu, E. Kim, M. Minamino, R.d. Groot, A.J. Katan, A. Bonato, D. Marenduzzo, D. Michieletto, F. Uhlmann, C. Dekker, Bridging-induced phase separation induced by cohesin SMC protein complexes, Science Advances, 7 (2021) eabe5905.

[26] F.M. Kelley, B. Favetta, R.M. Regy, J. Mittal, B.S. Schuster, Amphiphilic proteins coassemble into multiphasic condensates and act as biomolecular surfactants, Proceedings of the National Academy of Sciences, 118 (2021).

[27] S. Qamar, G. Wang, S.J. Randle, F.S. Ruggeri, J.A. Varela, J.Q. Lin, E.C. Phillips, A. Miyashita, D. Williams, F. Ströhl, W. Meadows, R. Ferry, V.J. Dardov, G.G. Tartaglia, L.A. Farrer, G.S. Kaminski Schierle, C.F. Kaminski, C.E. Holt, P.E. Fraser, G. Schmitt-Ulms, D. Klenerman, T. Knowles, M. Vendruscolo, P. St George-Hyslop, FUS Phase Separation Is Modulated by a Molecular Chaperone and Methylation of Arginine Cation-$\pi$ Interactions, Cell, 173 (2018) 720-734.e715.

[28] C. Shi, G. Du, J. Wang, P. Sun, T. Chen, Polyelectrolyte–Surfactant Mesomorphous Complex Templating: A Versatile Approach for Hierarchically Porous Materials, Langmuir, 36 (2020) 1851-1863.

[29] O. Santos, E.S. Johnson, T. Nylander, R.K. Panandiker, M.R. Sivik, L. Piculell, Surface Adsorption and Phase Separation of Oppositely Charged Polyion–Surfactant Ion Complexes: 3. Effects of Polyion Hydrophobicity, Langmuir, 26 (2010) 9357-9367.

[30] P. Linse, L. Piculell, P. Hansson, Models of Polymer-Surfactant Complexation, in: J.C.T. Kwak (Ed.) Polymer-Surfactant Systems, CRC Press, Madison Avenue, New York, USA, 1998.

[31] M. Ghasemi, S.N. Jamadagni, E.S. Johnson, R.G. Larson, A Molecular Thermodynamic Model of Coacervation in Solutions of Polycations and Oppositely Charged Micelles, Langmuir, 39 (2023) 10335-10351.

[32] M. Nguyen, K. Shen, N. Sherck, S. Köhler, R. Gupta, K.T. Delaney, M.S. Shell, G.H. Fredrickson, A molecularly informed field-theoretic study of the complexation of polycation PDADMA with mixed micelles of sodium dodecyl sulfate and ethoxylated surfactants, The European Physical Journal E, 46 (2023).

[33] J.J. Madinya, C.E. Sing, Hybrid Field Theory and Particle Simulation Model of Polyelectrolyte–Surfactant Coacervation, Macromolecules, 55 (2022) 2358-2373.

[34] E. Guzmán, L. Fernández-Peña, G. S. Luengo, A. Rubio, A. Rey, F. Léonforte, Self-Consistent Mean Field Calculations of Polyelectrolyte-Surfactant Mixtures in Solution and upon Adsorption onto Negatively Charged Surfaces, Polymers, 12 (2020) 624.

[35] M. Andersson, P. Hansson, Phase Behavior of Salt-Free Polyelectrolyte Gel–Surfactant Systems, The Journal of Physical Chemistry B, 121 (2017) 6064-6080.

[36] Y.D. Gordievskaya, A.M. Rumyantsev, E.Y. Kramarenko, Polymer gels with associating side chains and their interaction with surfactants, The Journal of Chemical Physics, 144 (2016).

[37] A.M. Rumyantsev, S. Santer, E.Y. Kramarenko, Theory of Collapse and Overcharging of a Polyelectrolyte Microgel Induced by an Oppositely Charged Surfactant, Macromolecules, 47 (2014) 5388-5399.

[38] P.S. Kuhn, A. Diehl, Flexible polyelectrolyte conformation in the presence of oppositely charged surfactants, Physical Review E, 76 (2007).

[39] R.J. Allen, P.B. Warren, Phase behaviour of oppositely charged polymer/surfactant mixtures, Europhysics Letters, 64 (2003) 468–474.

[40] P. Hansson, S. Schneider, B. Lindman, Phase Separation in Polyelectrolyte Gels Interacting with Surfactants of Opposite Charge, The Journal of Physical Chemistry B, 106 (2002) 9777-9793.

[41] P. Hansson, Self-Assembly of Ionic Surfactants in Polyelectrolyte Solutions: A Model for Mixtures of Opposite Charge, Langmuir 17 (2001) 4167-4180.

[42] A.J. Konop, R.H. Colby, Role of Condensed Counterions in the Thermodynamics of Surfactant Micelle Formation with and without Oppositely Charged Polyelectrolytes, Langmuir, 15 (1999) 58-65.

[43] K. Kogej, Polyelectrolytes and Surfactants in Aqueous Solutions: From Dilute to Concentrated Systems, in: A. Iglič (Ed.) Advances in Planar Lipid Bilayers and Liposomes, Academic Press, Cambridge, Massachusetts, United States, 2012, pp. 199-237.

[44] X. Dou, S. Gao, Z. Lu, J. Huang, Y. Yan, Effect of the Molecular Weight of Polyelectrolyte and Surfactant Chain Length on the Solid-Phase Molecular Self-Assembly, The Journal of Physical Chemistry B, 127 (2023) 10923-10930.

[45] A. Svensson, J. Norrman, L. Piculell, Phase Behavior of Polyion−Surfactant Ion Complex Salts: Effects of Surfactant Chain Length and Polyion Length, The Journal of Physical Chemistry B, 110 (2006) 10332–10340.

[46] K. Thalberg, B. Lindman, G. Karlstroem, Phase behavior of systems of cationic surfactant and anionic polyelectrolyte: influence of surfactant chain length and polyelectrolyte molecular weight, The Journal of Physical Chemistry, 95 (1991) 3370–3376.

[47] C. Yomota, Y. Ito, M. Nakagaki, Interaction of Cationic Surfactant with Arabate and

Chondroitin Sulfate, Chemical and Pharmaceutical Bulletin, 35 (1987) 798-807.

[48] M.L. Huggins, Solutions of Long Chain Compounds, The Journal of Chemical Physics, 9 (1941) 440-440.

[49] P.J. Flory, Thermodynamics of High Polymer Solutions, The Journal of Chemical Physics, 10 (1942) 51-61.

[50] M.T. Bashford, E.M. Woolley, Enthalpies of dilution of aqueous decyl-, dodecyl-, tetradecyl-, and hexadecyltrimethylammonium bromides at 10, 25, 40, and 55°C, The Journal of Physical Chemistry, 89 (1985) 3173–3179.

[51] G.A. Ahmadova, R.A. Rahimov, A.Z. Abilova, K.A. Huseynova, E. Imanov, F.I. Zubkov, Effect of head-group of cationic surfactants and structure of ionic groups of anionic polyelectrolyte in oppositely charged polymer-surfactant complexes, Colloids and Surfaces A: Physicochemical and Engineering Aspects, 613 (2021) 126075.

[52] P. Yan, C. Jin, C. Wang, J. Ye, J.-X. Xiao, Effect of surfactant head group size on polyelectrolyte–surfactant interactions: steady-state and time-resolved fluorescence study, Journal of Colloid and Interface Science, 282 (2005) 188-192.

[53] M. Muthukumar, Double screening in polyelectrolyte solutions: Limiting laws and crossover formulas, The Journal of Chemical Physics, 105 (1996) 5183-5199.

[54] M. Muthukumar, 50th Anniversary Perspective: A Perspective on Polyelectrolyte Solutions, Macromolecules, 50 (2017) 9528-9560.

[55] H. Yong, Reentrant condensation of polyelectrolytes induced by diluted multivalent salts: the role of electrostatic gluonic effect, Biomacromolecules, 25 (2024), DOI: 10.1021/acs.biomac.4c01037.

[56] J.P.d. Souza, H.A. Stone, Exact analytical solution of the Flory–Huggins model and extensions to multicomponent systems, The Journal of Chemical Physics, 161 (2024) 044902.

[57] D. Qian, T.C.T. Michaels, T.P.J. Knowles, Analytical Solution to the Flory–Huggins Model, The Journal of Physical Chemistry Letters, 13 (2022) 7853-7860.

[58] C.N. Pace, H. Fu, K.L. Fryar, J. Landua, S.R. Trevino, B.A. Shirley, M.M. Hendricks, S. Iimura, K. Gajiwala, J.M. Scholtz, G.R. Grimsley, Contribution of Hydrophobic Interactions to Protein Stability, Journal of Molecular Biology, 408 (2011) 514-528.

[59] U. Kim, W.M. Carty, Effect of polymer molecular weight on adsorption and suspension rheology, Journal of the Ceramic Society of Japan, 124 (2016) 484-488.

[60] A.P. Safronov, L.V. Adamova, A.S. Blokhina, I.A. Kamalov, P.A. Shabadrov, Flory-Huggins parameters for weakly crosslinked hydrogels of poly(acrylic acid) and poly(methacrylic acid) with various degrees of ionization, Polymer Science Series A, 57 (2015) 33-42.

[61] R.J. Baxter, Exactly Solved Models in Statistical Mechanics, Academic Press, London, England, 1982.

[62] H.E. Stanley, Introduction to Phase Transitions and Critical Phenomena (Reprint Edition), Oxford University Press, Oxford, England, 1987.

[63] H. Yong, J.-U. Sommer, Cononsolvency Effect: When the Hydrogen Bonding between a Polymer and a Cosolvent Matters, Macromolecules, 55 (2022) 11034-11050.

[64] H.-W. Tseng, P.-C. Chen, H.-W. Tsui, C.-H. Wang, T.-Y. Hu, L.-J. Chen, Effect of molecular weight of poly(acrylic acid) on the interaction of oppositely charged ionic surfactant–polyelectrolyte mixtures, Journal of the Taiwan Institute of Chemical Engineers, 92 (2018) 50-57.

[65] Y. Wang, K. Kimura, P.L. Dubin, W. Jaeger, Polyelectrolyte−Micelle Coacervation: Effects of Micelle Surface Charge Density, Polymer Molecular Weight, and Polymer/Surfactant Ratio, Macromolecules, 33 (2000) 3324–3331.

[66] Y. Li, J. Xia, P.L. Dubin, Complex Formation between Polyelectrolyte and Oppositely Charged Mixed Micelles: Static and Dynamic Light Scattering Study of the Effect of Polyelectrolyte Molecular Weight and Concentration, Macromolecules, 27 (1994) 7049–7055.

[67] J.-C. Lee, Rheological Properties of Polyelectrolyte and Thermodynamic Properties of Polyelectrolyte/Surfactant System at Different Degrees of Neutralization (open access), Department of Chemical Engineering, National Taiwan University, Taipei, 2024, pp. 1-96, DOI: 10.6342/NTU202403488.

[68] E. Spruijt, S.A.v.d. Berg, M.A.C. Stuart, J.v.d. Gucht, Direct Measurement of the Strength of Single Ionic Bonds between Hydrated Charges, ACS Nano, 6 (2012) 5297–5303.

[69] J.N. Israelachvili, Intermolecular and Surface Forces (3rd Edition), Academic Press, London, 2011.

[70] L. Sun, Q. Lei, B. Peng, G.M. Kontogeorgis, X. Liang, An analysis of the parameters in the Debye-Hückel theory, Fluid Phase Equilibria, 556 (2022) 113398.